\pdfoutput=1
\documentclass[12pt, reqno]{article}
\usepackage{epsfig}
\usepackage{amssymb}
\usepackage{amsmath}
\usepackage{mathrsfs}
\usepackage{cite}
\usepackage{hyperref}

\newcommand{\be}{\begin{eqnarray}}
\newcommand{\ee}{\end{eqnarray}}

\newcommand{\ab}[1]{\langle #1\rangle}

\begin{document}

\baselineskip=18pt

\setcounter{footnote}{0}
\setcounter{figure}{0}
\setcounter{table}{0}

\begin{titlepage}

\begin{center}

{\Large \bf A Note on Polytopes for Scattering Amplitudes}

\vspace{0.2cm}

{\bf N. Arkani-Hamed$^a$, J. Bourjaily $^{a,b}$, F. Cachazo$^{c}$, A. Hodges$^{d}$ and J. Trnka$^{a,b}$}

\vspace{.2cm}

{\it $^{a}$ School of Natural Sciences, Institute for Advanced Study, Princeton, NJ 08540, USA}

{\it $^{b}$ Department of Physics, Princeton Uniersity, Princeton, NJ 08544, USA}

{\it $^{c}$ Perimeter Institute for Theoretical Physics, Waterloo, Ontario N2J W29, CA}

{\it $^{d}$ Wadham College, University of Oxford, Oxford OX1 3PN, UK}

\end{center}

\begin{abstract}
In this note we continue the exploration of the polytope picture for scattering amplitudes, where amplitudes are associated with the volumes of  polytopes in generalized momentum-twistor spaces. After a quick warm-up example illustrating the essential ideas with the elementary geometry of polygons in ${\mathbb{CP}}^2$, we interpret the 1-loop MHV integrand as the volume of a polytope in ${\mathbb{CP}}^3 \times {\mathbb{CP}}^3$, which can be thought of as the space obtained by taking the geometric dual of the Wilson loop in each ${\mathbb{CP}}^3$ of the product. We then review the polytope picture for the  NMHV tree amplitude and give it a more direct and intrinsic definition  as the geometric dual of a canonical ``square" of the Wilson-Loop polygon, living in a certain extension of momentum-twistor space into ${\mathbb{CP}}^4$. In both cases, one natural class of triangulations of the polytope produces the BCFW/CSW representations of the amplitudes; another class of triangulations leads to a striking new form, which is both remarkably simple as well as manifestly cyclic and local.

\end{abstract}

\bigskip
\bigskip

\end{titlepage}

\section{Towards a Geometry of Scattering Amplitudes}

Recent months have seen significant advances in our understanding of perturbative scattering amplitudes in gauge theories, especially for ${\cal N} = 4$ SYM in the planar limit.
A generalization of the BCFW recursion relations \cite{Britto:2004ap,Britto:2005fq} to all loops has been given  to determine the planar integrand \cite{ArkaniHamed:2010kv} of the theory, naturally formulated in momentum-twistor space \cite{Hodges:2009hk}, making the Yangian symmetry \cite{Drummond:2009fd} manifest, and  extending the Grassmannian duality for leading singularities \cite{ArkaniHamed:2009dn} to the full theory.  The integrand has also been beautifully interpreted \cite{Mason:2010yk,CaronHuot:2010ek} as a
supersymmetric generalization of the null-polygonal Wilson-Loop \cite{Alday:2007hr}, making dual-superconformal invariance \cite{Drummond:2006rz, Alday:2007hr,Berkovits:2008ic, Drummond:2008vq} manifest and providing a general proof
\cite{CaronHuot:2010ek,Bullimore:2010dz} of the Wilson-Loop/Amplitude duality \cite{Alday:2007hr}.

Despite these advances, our understanding of the integrand still leaves something to be desired. The definition in terms of either scattering amplitudes or the Wilson-Loop only manifests half of the superconformal symmetries of the theory, obscuring the infinite-dimensional Yangian symmetry; it also
invokes gauge redundancies that are made necessary by any local Lagrangian description. The BCFW representation of the amplitude is more compact, and gives a complete definition of the theory making no direct reference to  space-time notions. However it is not manifestly cyclically invariant: there are many different BCFW forms, depending on the choice of legs for BCFW deformation. All of this suggests that the various formulations for scattering amplitudes that have been uncovered so far are different representations of a single underlying object, which awaits a deeper, more intrinsic and invariant characterization.

In this brief note we take some preliminary steps towards uncovering this underlying structure. We will study the simplest non-trivial amplitudes in the theory--the tree-level NMHV amplitudes, and the integrand for the 1-loop MHV amplitudes. Following and generalizing the observations of \cite{Hodges:2009hk}, we will interpret these amplitudes as the volumes of polytopes in certain extensions of momentum-twistor space. To actually evaluate the volume, we need to triangulate the polytope into elementary simplices, and one natural choice of triangulation leads directly to BCFW (and CSW) representations of these amplitudes. However we also find even simpler triangulations of the polytopes that yield completely new expressions for these familiar old friends.

The BCFW representation of the NMHV tree amplitude, written in momentum-twistor space, is
\be
\label{NMHV}
M^{{\rm NMHV}}_n = \frac{1}{2}\sum_{i,j} \left[1 \, i \, i+1 \, j \, j+1\right]
\ee
where
\be
\left[abcde\right] = \frac{\delta^4\left(\eta_a \ab{bcde} + \cdots + \eta_e \ab{abcd}\right)}{\ab{bcde} \ab{cdea} \ab{deab} \ab{eabc} \ab{abcd}}
\ee
is the basic ``$R$-invariant" \cite{Drummond:2008vq} written in momentum-twistor space \cite{Mason:2009qx}.
Similarly, the 1-loop integrand for MHV amplitudes is \cite{ArkaniHamed:2010kv}
\be
\label{MHV}
M^{{\rm MHV}}_{{\rm 1-loop}, n} = \frac{1}{2}\sum_{i,j} \left[1 \, i \, i+1 ; 1 \, j \, j+1 \right]
\ee
where we have introduced objects $\left[a b c ; x y z\right]$ via
\be
\left[a b c; x y z\right] = \frac{\left(\ab{A a b c} \ab{B x y z} - \ab{A x y z} \ab{B a b c}\right)^2}{\ab{AB ab} \ab{AB bc} \ab{AB ca} \ab{AB xy} \ab{AB yz} \ab{AB zx}}.
\ee

Note the presence, in both of these formulas, of the special point ``1". The CSW representation \cite{ArkaniHamed:2009sx,Bullimore:2010dz} of the same amplitude is obtained by replacing ``1" in this formula with a general momentum twistor, $Z_1 \to Z_*$.

The BCFW/CSW expressions for the amplitudes are not manifestly cyclically invariant/independent of the auxiliary twistor $Z_*$, nor are they manifestly free of spurious poles. All these properties only emerge after the summation is performed.
There is a nice algebraic way of seeing this. The basic objects appearing in the formulas satisfy identities that allow us to express them in different
ways. In general, such identities follow from Grassmannian residue theorems \cite{ArkaniHamed:2009dn}. We can also understand them in the following simple way. Take any Yangian invariant object $Y_{n,k}(Z_1, \cdots, Z_n)$, which a residue of the Grassmannian integral. Consider its BCFW deformation under $Z_n \to Z_n + z Z_{n-1}$. It is easy to show that the residues of all the poles in the complex $z$ plane are {\it also} Grassmannian residues, and are therefore Yangian invariant. Thus, an application of Cauchy's theorem on the BCFW deformed function of $z$ yields a relation between Yangian invariants.

Starting with the basic NMHV $R$-invariant $[a b c d e]$, we can slightly generalize this procedure and consider a general deformation $Z_a \to Z_a + z Z_f$. Then, Cauchy's theorem yields the familiar 6-term identity
\be
\label{6term}
\left[abcde\right] + \left[bcdef\right] + \left[cdefa\right] + \left[defab\right] + \left[efabc\right] + \left[fabcd\right] = 0 .
\ee
We can also apply this to the basic objects $[a b c; x y z]$. It is interesting to note here that that while the $[1 \, i \, i+1; 1 \, j\, j+1]$ are indeed Yangian invariant, the general $[a b c; x y z]$ are not. Nonetheless under the deforming $Z_a \to Z_a + z Z_d$, or $Z_x \to Z_x + z Z_w$, Cauchy's theorem yields 4-term identities
\begin{eqnarray}
\left[abc;xyz\right] + \left[bcd;xyz\right] + \left[cda;xyz\right] + \left[dab;xyz\right] &=& 0, \nonumber \\
\left[abc;xyz\right] + \left[abc;yzw\right] + \left[abc;zwx\right] + \left[abc;wxy\right] &=& 0.
\end{eqnarray}

Suppose we are given some linear combination of the $\left[abcde\right]$'s. Given these identities, we can have two different linear combinations representing the same function. How then can we determine if two expressions are equal? More formally, how can we characterize the equivalence class of linear combinations of $R$-invariants, which differ by these identities?

The key is to note that the $R$-invariant identity can formally be written using a ``boundary" operation. Imagine what is (for now) a completely formal object, a ``5-simplex" $[abcdef]$, which is completely antisymmetric in its indices. Then, the linear combination of $R$-invariants entering the 6-term identity is just the ``boundary" of this simplex, and the identity becomes
\be
\partial \left[abcdef\right] = 0.
\ee
Now suppose that $\alpha$ and $\beta$ are two linear combinations of $R$-invariants. We wish to determine if $\alpha = \beta$ up to 6-term identities; that is we want to determine whether there exists some simplex $\sigma$ such that
\be
\alpha = \beta + \partial \sigma.
\ee
Since $\partial^2 = 0$, it suffices to check that
\be
\partial \alpha = \partial \beta.
\ee
Thus we have learned that while any given representation of an amplitude in terms of a sum of $R$-invariants is not unique, the ``boundary" of the amplitude {\it is} invariant. Using the standard definition of the boundary operation on simplices, the ``boundary" of one of the $R$-invariants is
\be
\partial \left[a b c d e \right] = \left[a  b  c d \right] + \left[b  c  d e \right] +
\left[c d  e a \right] + \left[d  e  a b \right] + \left[e  a  b c \right].
\ee
Note that the boundary is also a list of the poles occurring the definition of the $R$-invariant--this is not an accident, as will become clear in the polytope picture of the next sections.

We can now easily compute the boundary of the BCFW/CSW forms of the NMHV tree amplitude:
\be
\partial \sum_{i,j} \left[* \, i \, i+1 \, j \, j+1 \right]  = \sum_{i,j} \left[i \, i+1 \, j \, j+1 \right]
\ee
which is a beautifully cyclic object, independent of the point $Z_*$, in one-to-one correspondence with the democratic sum over all the physical poles of the NMHV amplitude! This is enough to prove that the BCFW/CSW forms of the NMHV amplitude define a cyclic object free of spurious poles. It is possible to prove something stronger: the {\it only} combination of $R$-invariants that is free of spurious poles is the NMHV tree amplitude! This is because if the amplitude is free of spurious poles, its boundary must only contain ``physical" 3-simplices of the form $[i \,  i+1 \, j \, j+1]$. Since this is supposed to be a boundary, {\it its} boundary must vanish. It is easy to see that the only combination of physical 3-simplices that it boundary free is $\sum_{i,j} [i \, i+1 \, j \, j+1]$.

We can do exactly the same exercise for the MHV 1-loop amplitude. The 4-term identities can be interpreted as 
\be
[\partial (abcd), xyx] = 0, \quad [abc, \partial (xyzw)] = 0. 
\ee
This makes it natural to define a boundary operation on $\left[\sigma_1;\sigma_2\right]$ as acting separately on the two simplices $\sigma_1, \sigma_2$,
\be
\partial \left[\sigma_1;\sigma_2\right] = \left[\partial \sigma_1; \partial \sigma_2\right].
\ee
The boundary of the individual terms in \mbox{eqn. (\ref{MHV})} are again in one-to-one correspondence with their poles.
The boundary of the BCFW/CSW form of the 1-loop amplitude is
\be
\partial \sum_{i,j} \left[* \, i \, i+1; * \, j \, j+1\right] = \sum_{i,j} \left[i \, i+1; j \, j+1\right].
\ee
Again this is a beautifully cyclic object, independent of the point $Z_*$,  in one-to-one correspondence with all the physical poles of the integrand. We can in fact give a slightly more general expression for the MHV integrand using {\it two} reference twistors $Z_*,Z_{*^\prime}$, which has the same boundary, as
\be
M^{{\rm MHV}}_{{\rm 1-loop},n} = \frac{1}{2}\sum_{i,j} \left[* \, i \, i+1; *^\prime \, j \, j+1 \right].
\ee
There are of course many other representations of these amplitudes; again, only the ``boundary" is invariant.

We have gone through this discussion in some detail because these sorts of algebraic arguments --- reflecting Grassmannian residue theorems --- generalize readily to more complicated amplitudes \cite{wip}. In the rest of this note, however, we will pursue a different line of thought. Taking our cue from \cite{Hodges:2009hk}, we will see that the appearance of ``simplices" and ``boundaries" is not an accident, but has a deeper geometric origin. As we have mentioned already, the amplitudes will be interpreted as the volumes of certain polytopes. Certain triangulations of these polytopes give a very pretty and direct geometric interpretation of the BCFW/CSW representations of the amplitudes. But the picture does more than simply re-organize algebraic manipulations in geometric terms: a different triangulation of the polytopes leads to entirely new representations of the amplitudes, which are both strikingly simple and manifestly cyclic and {\it local}.

We will begin our discussions with an extremely simple warm-up exercise familiar from elementary plane geometry. We then move on to discussing the MHV 1-loop and NMHV tree amplitudes.

\section{Warm-Up: The Area of Polygons in ${\mathbb{CP}}^2$}

Consider a simple set of functions
\be
[abc] = \frac{1}{2} \frac{\ab{abc}^2}{\ab{Aab} \ab{Abc} \ab{Aca}}.
\ee
Here $Z_{a,b,c}$ and $Z_A$  are in $\mathbb{C}^3$. Since $[abc]$ has weight zero under rescaling $a,b,c$ we can also think of $Z_{a,b,c}$ as points in $\mathbb{CP}^2$, corresponding to twistors for our ``external particles", while $Z_A$ is a reference twistor.

Clearly the $[abc]$'s are analogous to (a particular Grassmann component of) the $R$-invariants $[abcde]$ we are familiar with, and are also very closely related to the $[abc;xyz]$ objects appearing in the 1-loop MHV amplitude. Let us define an analog of the ``amplitude", $A_n$, via a ``BCFW/CSW" formula of the form
\be
A_n = \sum_i \left[* \, i \, i+1 \right].
\ee
This expression is manifestly cyclic invariant but term-by-term has ``unphysical poles" (where ``physical poles" are only of the form $\ab{A \,j\, j+1}$). However, we can easily see that the sum is in fact cyclic and only has physical poles. Following our earlier discussion, we can derive identities satisfied by $[a b c]$ 
by deforming $Z_a \to Z_a + z Z_d$, to find a 4-term identity
\be
[abc] + [bcd] + [cda] + [dab] = 0.
\ee
We can think of this formally as
\be
\partial [abcd] = 0.
\ee
Following our earlier logic, we can define the ``boundary" of $[abc]$ itself as
\be
\partial [abc] = [ab] + [bc] + [ca]
\ee
which is in one-to-one correspondence with the poles in $[abc]$. Finally, we can compute the ``boundary" of the ``amplitude" to find
\be
\partial A_n = \sum_i \left[i \, i+1 \right]
\ee
which is just the democratic sum over all ``physical" boundaries.

These observations make it natural to associate $[abc]$ with a triangle in $\mathbb{CP}^2$ whose vertices are $Z_a,Z_b,Z_c$, and the amplitude itself with the interior of the polygon $L_n$ with vertices $Z_i$ and edges $(Z_i, Z_{i+1})$, as in the figure below for the case of six particles:
\be
\includegraphics[scale=.85]{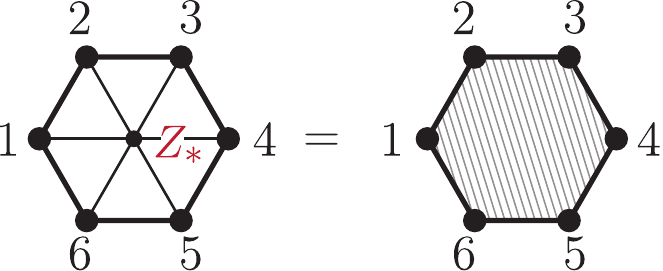} \nonumber
\ee

Now the $[abc]$ certainly have the same additive structure as the simplices defined by the triangles $(abc)$. We should therefore be able to give a formula for $[abc]$ as a function of the triangle $(abc)$, in a way that preserves this additive structure. This is very easy to do. The function $[abc]$ is the {\it area} of the {\it geometric dual} triangle to $[abc]$ in $\mathbb{CP}^2$, whose {\it edges} are the dual lines to $Z_a,Z_b,Z_c$:
\be
\includegraphics[scale=.75]{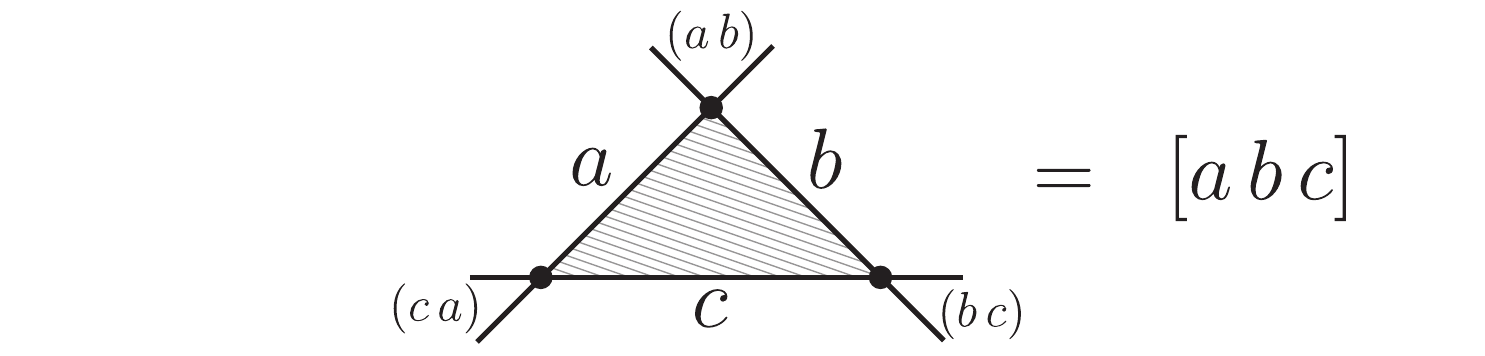} \nonumber
\ee
The amplitude is then simply the area of the geometric dual  $\widetilde{L}_n$ of the polygon $L_n$:
\be
\includegraphics[scale=.65]{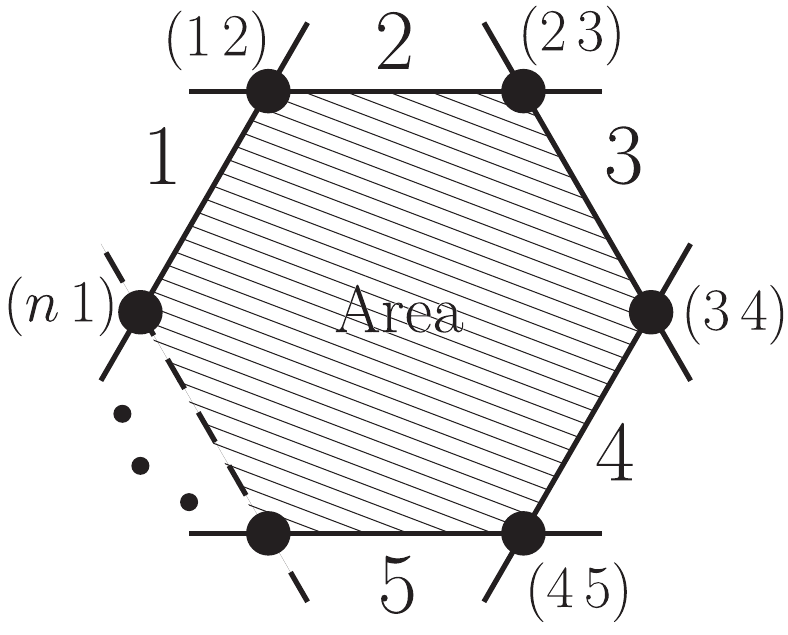} \nonumber
\ee
Let us see how this works explicitly by doing some very elementary plane geometry.  Let the twistors $Z^I_{a,b,c,\cdots}$ and the reference twistor $Z_A^I$ have an upstairs SL(3) index. We are interested in the dual space whose co-ordinates $W_I$ have a lower SL(3) index. Now, suppose we are given three points $W^{1}_I, W^{2}_I, W^{3}_I$. As is standard in projective geometry,  the point $Z^I_A$ breaks SL(3) but leaves an SL(2) invariant, and defines a projection direction. Putting
\be
Z_A^I = \left( \begin{array}{c} 0\\ 0 \\ 1 \end{array} \right), \, \, W_I = \left( \begin{array}{c} x \\ y \\ 1 \end{array} \right)
\ee
we can think of the points $(x,y)$ as lying in a two-dimensional plane, on which the unbroken SL(2) acts. The area of the triangle associated with $W^{1},W^{2},W^{3}$ is the SL(2) invariant given by
\be
{\rm Area}(W^{1}, W^{2}, W^{3}) =\frac{1}{2} \left| \begin{array}{ccc} x^1 & x^2 & x^3 \\ y^1 & y^2 & y^3 \\ 1 & 1 & 1\end{array} \right|
\ee
which we can write in a projectively invariant way as
\be
\label{Area}
{\rm Area}(W^{1} W^{2} W^{3}) = \frac{1}{2}
\frac{\ab{W^{1} W^{2} W^{3}}}{(Z_A \cdot W^{1})(Z_A \cdot W^{2})(Z_A \cdot W^{3})}.
\ee
Note that this is {\it not} invariant under rescaling the reference twistor $Z_A$, which is appropriate, since $Z_A$ defines the plane in which the area is defined and the area is not dimensionless.

Now, suppose we are given instead three points in the original space, $Z_a^I,Z_b^I,Z_c^I$. Each of these points is associated with a line in the $W$ space, with e.g. the point $a$ defining the line $Z_a^I W_I = 0$. The lines $a$ and $b$ intersect at the point $(ab)$ in $W$ space, with co-ordinate $W^{(ab)}_I = \epsilon_{IJK} Z_a^J Z_b^K$. Thus, the area of this dual triangle is
\be
{\rm Area}(\widetilde{[abc]}) = \frac{1}{2} \frac{\ab{(ab) (bc) (ca)}}{\ab{A ab} \ab{A bc} \ab{A ca}} =
\frac{1}{2} \frac{\ab{abc}^2}{\ab{Aab} \ab{Abc} \ab{Aca}} = [abc].
\ee

With these elementary facts in hand, it is easy to identify the triangulations of the polygon associated with the BCFW/CSW representations of the amplitude, which correspond to triangulating  $\widetilde{L}_n$, with the dual triangles $\widetilde{[* \,i \, i+1]}$. An example of a BCFW triangulation for the 4-particle amplitude is shown below:
\be
\includegraphics[scale=.75]{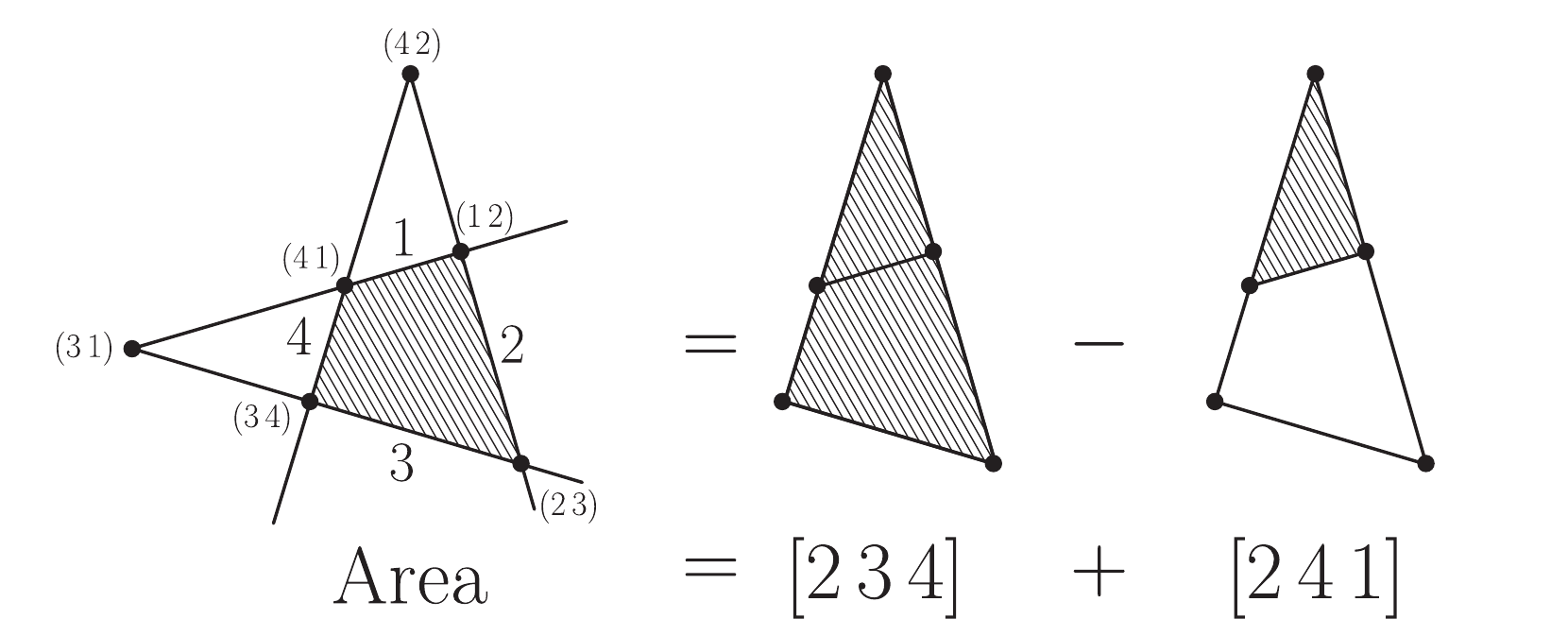} \nonumber
\ee
Note that the BCFW triangulation is characterized by not introducing any new {\it lines}, but certainly introduces new vertices. However, we have an even more obvious triangulation of the same object, introducing a dual reference point $W_*$, and triangulating directly using the vertices as
\be
\includegraphics[scale=.75]{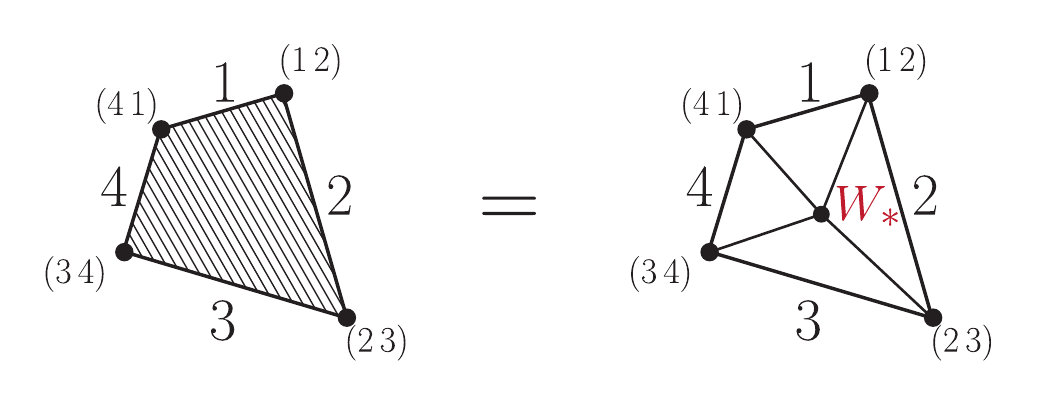} \nonumber
\ee
For a general polygon, the area can be triangulated using the triangles with vertices
$(W_*, \, (i-1\, i), \, (i\, i+1))$. We can compute the area of this triangle using \mbox{eqn. (\ref{Area})}, giving an $n-$term expression for the amplitude
\be
\label{local}
A_n = \frac{1}{2} \sum_i \frac{\ab{W_* \, (i-1 \,i) \, (i\, i+1)}}{Z_A \cdot W_* \ab{A \, i-1\, i} \ab{A \, i \, i+1}} = \sum_i \frac{(Z_i \cdot W_*) \ab{i-1 \, i \, i+1}}{(Z_A \cdot W_*) \ab{A\, i-1\, i} \ab{A\, i\, i+1}}.
\ee

Note that in this form, {\it all} the poles involving the $Z_i$ are manifestly ``physical". This is obvious, since we have triangulated the polygon only using its vertices, and the divergences of the amplitude can only occur if some vertex $(k\, k+1)$ moves off to infinity, making the area diverge. By contrast, the BCFW/CSW representations introduce new points in $W$ space, with associated spurious pole which cancel in the sum.
Note also that this triangulation involved a natural reference point $W_*$, analogous to the reference point $Z_*$ in the CSW representation of the amplitude. The result is independent of $W_*$, but term-by-term has a ``spurious pole" $(Z_A \cdot W_*)$. We can choose $W_*$ to coincide with one of the external points, say $W_* = (k \, k\!+\!1)$, giving an $(n-2)$ term expression with manifestly physical poles which is however no longer manifestly cyclic invariant.

We close this warm-up section with a few comments. We have drawn pictures of our polygons on a real 2-dimensional plane, but of course the functions are all holomorphic and defined on ${\mathbb{CP}}^2$. The complex areas have a very nice interpretation in terms of contour integrals in ${\mathbb{CP}}^2$ with boundaries on the polygon $\widetilde{L}_n$
\cite{Hodges:2009hk}. It is perhaps easiest to get a feeling for such contour integrals with boundary by considering the simplest case of a standard integral over one complex variable $z$, thought of as a projective integral over ${\mathbb{CP}}^1$. Let's introduce a variable $w_I = (x,y)$ in ${\mathbb{C}}^2$. Consider an integral with boundaries on $z_a^I w_I = 0, z_b^I w_I = 0$:
\be
\int \limits_{\begin{array}{c} z_a \cdot w = 0 \\ z_b \cdot w = 0 \end{array}} \frac{Dw}{(z_c \cdot w)^2}
\ee
here, we use
\be
D^{n-1} w \equiv \frac{d^n w}{{\rm vol GL(1)}}
\ee
to denote the measure on ${\mathbb{CP}}^{n-1}$.
Using inhomogeneous co-ordinates $w = (1,z)$, the boundaries are at $w = -x_a/y_a$ and $w = -x_b/y_b$. The integrand is simply $\frac{1}{(x_c + y_c z)^2}$, and can be trivially integrated on any contour between the two end-points. The result is
\be
\int \limits_{\begin{array}{c} z_a \cdot w = 0 \\ z_b \cdot w = 0 \end{array}} \frac{D w}{(z_c \cdot w)^2} = \frac{\ab{ab}}{\ab{ac} \ab{cb}}.
\ee
This simple result generalizes readily to $\mathbb{CP}^n$; for instance our elementary triangle $[abc]$ is given by
\be
\left[abc\right] = \int \limits_{\begin{array}{c} Z_a \cdot W = 0 \\ Z_b \cdot W = 0 \\ Z_c \cdot W = 0 \end{array}} \frac{D^2 W}{(Z_A \cdot W)^3}.
\ee
This representation of the $[abc]$'s as a contour integral makes the additive properties and the 4-term identities manifest, since $Z_a,Z_b,Z_c$ define the boundaries of the integration region.
The full amplitude $A_n$ is expressed as a contour integral with boundaries given on the dual polygon $\widetilde{L}_n$:
\be
A_n = \int \limits_{\widetilde{L}_n} \frac{D^2 W}{(Z_A \cdot W)^3}.
\ee

In the following sections, we will need the generalization of the simple formula for the area of triangles in ${\mathbb{CP}}^2$ to the general volume of $(n-1)$-simplices in ${\mathbb{CP}}^{n-1}$, again projected along some specific direction $Z_A$. Specifying the vertices $W_I^{1}, \cdots, W_I^{n}$, the volume is obviously
\be
{\rm Vol}\left[W^1_I, \cdots, W^n_I\right] =\frac{1}{(n\!-\!1)!} \frac{\ab{W^{1} \cdots W^{n}}}{(Z_A \cdot W^{1}) \cdots (Z_A \cdot W^{n})}.
\ee
If instead we specify the $(n-1)$-simplex by giving its faces $Z_1^I, \cdots, Z_n^I$, the volume is
\be
{\rm Vol}\left[Z_1^I, \cdots, Z_n^I \right] = \frac{1}{(n\!-\!1)!}
 \frac{\ab{Z_1 \cdots Z_n}^n}{\ab{Z_A Z_1 \cdots Z_{n-1}} \ab{Z_A Z_2 \cdots Z_n} \cdots \ab{Z_A Z_n \cdots Z_{n-2}}}.
\ee

Finally, in our discussion of MHV 1-loop amplitudes, we will encounter plane polygons defined by twistors in ${\mathbb{CP}}^3$. Suppose we are given two reference twistors, $Z_A^I,Z_B^I$. The $Z_B^I$ define a plane and thus a ${\mathbb{CP}}^2$ inside the ${\mathbb{CP}}^3$. Restricting all the twistors to this plane, we can then project along the direction $Z_A$ to define the area as we did above.
Thus, given three twistors $Z_a^I,Z_b^I,Z_c^I$ in ${\mathbb{CP}}^3$, there is an associated area we can label $[abc]_{B;A}$, where the subscript reminds us that we are in the plane defined by $Z_B$ and projecting along $Z_A$. This area is given by
\be
\left[a b c\right]_{B;A} = \frac{\ab{B a b c}^2}{\ab{BAab} \ab{BAbc} \ab{BAca}}.
\ee
The area of the corresponding $n-$gon is
\be
A_{n \, B;A} = \sum_i \left[* \,i \, i+1 \right]_{B;A}.
\ee
We can also interpret this as a contour integral as
\be
A_{n \, B;A} = \int \frac{D^3W}{(Z_B \cdot W) (Z_A \cdot W)^3}
\ee
where the contour of integration is ``$S^1 \times$ Polygon", where the $S^1$ is evaluated around the pole $Z_B \cdot W = 0$, restricting the integral to the appropriate ${\mathbb{CP}}^2$, leaving us with the remaining boundary on the polygon $\widetilde{L}_n$.

We can also give a ``local" triangulation of $A_{n \, B;A}$ analogous to equation(\ref{local}). The reference point $W_{*}$ in ${\mathbb{CP}}^2$ can be obtained by  restricting a general reference bi-twistor $X$ to the $Z_B$ plane, via
$W_{* I} = \epsilon_{IJKL} Z_B^J X^{KL}$. This gives us
\be
\label{ABloc}
A_{n \, B;A} = \sum_i \frac{\ab{BX i}\ab{Bi-1 \, i \, i+1}}{\ab{BA X} \ab{BA i-1 \, i} \ab{BA i\, i+1}} .
\ee
There are obviously $n$ terms in this sum. Note again that all the poles involving the $Z_i$ are manifestly local. Each term does have a spurious pole $\ab{BAX}$, but of course these poles all cancel as the result is independent of $X$. We could make a special choice where $X = (k \, k+1)$
co-incides with one of the vertices of the polygon. This gives us an expression with only $(n-2)$ terms and no spurious poles of any kind, which is however not manifestly cyclically invariant.

\section{The One Loop MHV Integrand}

We now give a simple polytope interpretation of the 1-loop MHV integrand. Almost all the results we need were already discussed in our warm-up. Let us again examine the 1-loop MHV integrand in BCFW/CSW form
\be
\sum_{i,j} \frac{\left(\ab{A \, * \, i \, i+1} \ab{B \, * \, j \, j+1} - \ab{A \, * \, j \, j+1} \ab{B \, * \, i \, i+1}\right)^2}{\ab{AB \,* \, i} \ab{AB \,i \,i+1} \ab{AB \,i+1 \,*} \ab{AB \,* \,j} \ab{AB \,j \,j+1} \ab{AB \,j+1 \, *}} .
\ee
We will expand the square in the numerator. The first term is given by the sum
\begin{eqnarray}
\sum_i \frac{\ab{A \, * \, i \, i+1}^2}{\ab{AB \,* \, i} \ab{AB \,i \,i+1} \ab{AB \,i+1 \,*}} &\times& \sum_j \frac{\ab{B \, * \, j \, j+1}^2}{\ab{BA \,* \, j} \ab{BA \,j \,j+1} \ab{BA \, j+1 \,*}}.
\end{eqnarray}
We can recognize these expressions as computing the area $A_{n \, B;A}$ discussed in the warm-up section, corresponding to the area of the polygon restricted to the ${\mathbb{CP}}^2$ defined by $Z_B$ and projected along $Z_A$. The above sum becomes
\begin{eqnarray}
\sum_i \left[* \, i \, i+1\right]_{A;B} &\times& \sum_j \left[* \, j \, j+1\right]_{B;A} \nonumber \\ = \, A_{n \, A;B} &\times& A_{n \, B;A}.
\end{eqnarray}
Let's now look at the cross-term, which has the form
\begin{eqnarray}
-2 \sum_{i,j} \frac{\ab{A \, * \, i \, i+1} \ab{B \, * i \, i+1}}{\ab{AB \,* \, i} \ab{AB \,i \,i+1} \ab{AB \,i+1 \,*}} &\times& \frac{\ab{B \, * \, j \, j+1} \ab{A \, * \, j \, j+1}}{\ab{BA \,* \, j} \ab{BA \,j \,j+1} \ab{BA \, j+1 \,*}}.
\end{eqnarray}
We can also relate this to a polygon areas by using a differential operator 
\be
\frac{\ab{A \, * \, i \, i+1} \ab{B \, * i \, i+1}}{\ab{AB \,* \, i} \ab{AB \,i \,i+1} \ab{AB \,i+1 \,*}} = \frac{1}{2} \left(Z_B \cdot \frac{\partial}{\partial Z_A}\right) \frac{\ab{A \, * \, i \, i+1}^2}{\ab{AB \,* \, i} \ab{AB \,i \,i+1} \ab{AB \,i+1 \,*}}.
\ee
The cross-term becomes
\begin{eqnarray}
-\frac{1}{2} \, \left(Z_B \cdot \frac{\partial}{\partial Z_A}\right) \sum_i \left[* \, i \, i+1 \right] & \times &
\left(Z_A \cdot \frac{\partial}{\partial Z_B}\right) \sum_j \left[* \, j \, j+1 \right] \nonumber \\
= \, -\frac{1}{2} \, \left(Z_B \cdot \frac{\partial}{\partial Z_A}\right) A_{n \, A;B} & \times &
\left(Z_A \cdot \frac{\partial}{\partial Z_B}\right) A_{n \, B;A}.
\end{eqnarray}
We finally have
\be
\label{MHVPoly}
M^{{\rm MHV}}_{{\rm 1-loop}, n}= A_{n \, A;B} + A_{n \, B;A} - \frac{1}{2} \left(Z_B \cdot \frac{\partial}{\partial Z_A}\right) A_{n \, A;B} \times
\left(Z_A \cdot \frac{\partial}{\partial Z_B}\right) A_{n \, B;A}.
\ee
This expression of course makes the cyclic invariance of the integrand completely manifest.

We can interpret the MHV 1-loop integrand as a contour integral in a number of ways. The direct transcription of the expressions we have given is a contour integral of the form
\be
\int K(U,V,A,B) D^3 U D^3 V
\ee
where
\be
K(U,V,A,B) = \frac{1}{(B \cdot U)^3 (A \cdot V)} - \frac{2}{(B \cdot U)^2 (A \cdot V)^2} + \frac{1}{(B \cdot U) (A \cdot V)^3} .
\ee
The contour of integration is ``Polygon $\times S^1 \times S^1 \times$ Polygon" in the obvious way. This integral representation can be directly derived from a Fourier transformation of the Grassmannian formula for the MHV 1-loop integrand given in \cite{ArkaniHamed:2010kv}, but we won't pursue this interpretation further in this note.

Note that this form does not make it completely obvious that the integral depends on $A,B$ only through the line $(AB)$, though we can note that $Z_A \cdot \partial/\partial Z_B K =  Z_B \cdot \partial/\partial Z_A K = 0$. There is a more elegant representation as a contour integral, which makes the dependence on the line $(AB)$ more explicit:
\be
\int \frac{D^3U D^3 V}{\ab{UV AB}^4}
\ee
where the contour is ``Polygon $\times S^2 \times$ Polygon". Note that the integrand is not only explicitly only a function of the line $(AB)$, it is also only a function of the line $(UV)$; the integral over the $S^2$ leaves us with an integral over the Grassmannian $G(2,4)$.

Finally, returning to \mbox{eqn. (\ref{MHVPoly})}, we can obtain a local form of the MHV 1-loop integrand using the ``local triangulation" of $A_{n \, A;B}$ given in \mbox{eqn. (\ref{ABloc})}. It is natural to use two different reference bi-twistors $X,Y$ for triangulating $A_{n \, A;B}$ and $A_{n \, B;A}$. A short computation yields
\be
\label{localMHV}
M^{{\rm MHV}}_{{\rm 1-loop}, n} = \sum_{i,j} \frac{\ab{AB (Xi) \cap (Yj)} \ab{AB (i-1 \, i \, i+1) \cap(j-1 \, j \, j+1)}}{\ab{ABX} \ab{ABY} \ab{AB i-1 \, i} \ab{AB i \, i+1} \ab{AB j-1 \, j} \ab{AB j \, j+1}}.
\ee
Here
\begin{eqnarray}
\ab{AB (Xi) \cap (Yj)} &=& \ab{A X i} \ab{B Y j} - (A \leftrightarrow B), \nonumber \\ \ab{AB (i-1 \, i \, i+1) \cap (j-1 \, j \, j+1)} &= &\ab{A i-1 \, i \, i+1} \ab{B j-1 \, j \, j+1} - (A \leftrightarrow B).
\end{eqnarray}
This expression for the amplitude is also manifestly cyclic. Note that, in complete parallel with the discussion around \mbox{eqn. (\ref{ABloc})}, all the poles involving the $Z_i$ twistors are manifestly local. Each term has a dependence on the $X,Y$ bi-twistors, but these cancel in the sum which is independent of $X,Y$.
There are a number of obvious special cases of interests for this new form of the integrand. For instance we can take the two bi-twistors $X$ and $Y$ to be equal, yielding the form
\be
\sum_{i,j} \frac{\ab{X i j} \ab{AB (i-1 \, i \, i+1) \cap(j-1 \, j \, j+1)}}{\ab{ABX}\ab{AB i-1 \, i} \ab{AB i \, i+1} \ab{AB j-1 \, j} \ab{AB j \, j+1}}.
\ee
We can further take $X$ to be a simple bi-twistor corresponding to one of the external points, for instance $X = (n1)$.
With this choice there are no spurious poles of any sort, but the result is not manifestly cyclic invariant. Averaging over all cyclic images yields the local form of the integrand given in \cite{ArkaniHamed:2010kv}.

The equality between the local form \mbox{eqn. (\ref{localMHV})} and the BCFW form of \mbox{eqn. (\ref{MHV})} is a highly non-trivial identity between rational functions that we have now understood geometrically. As stressed in \cite{ArkaniHamed:2010kv}, the loop integrand is a well-defined object in the planar limit of any theory, and should manifest all symmetries. While the BCFW form on the integrand  exhibits the Yangian invariance of the theory, the local form is also crucially needed
\cite{ArkaniHamed:2010kv} for a physical IR  regularization \cite{Alday:2009zm,Hodges:2010kq,Mason:2010pg} of the theory. It is therefore very pleasing to see the two forms related in such a direct way.

\section{The NMHV Tree Amplitude}
We now turn to the NMHV tree amplitude, which has been given a polytope volume interpretation in \cite{Hodges:2009hk}. In \cite{Hodges:2009hk}, the description of the polytope was closely associated with the BCFW representation of the amplitude. We will begin by describing the polytope in a slightly more invariant way.

As a consequence of the 6-term identity \mbox{eqn. (\ref{6term})} for $R$-invariants, we can observe that the $\left[abcde\right]$'s have exactly the same additive properties as 4-simplices in $\mathbb{CP}^4$. In our warm-up example, we associated the $[abc]$'s with the area of a dual triangle in ${\mathbb{CP}}^2$. We would like to do the same for the $R$-invariants $\left[abcde\right]$, but there is a small difference: each particle $i$ is labeled by a supertwistor $(Z^I_i,\eta^\alpha_i)$, where $\alpha = 1, \cdots, 4$ is the SU(4) R-symmetry index. In order to proceed we have to associate a point in ${\mathbb{CP}}^4$ with this super-twistor. Fortunately this is easy to do. Let's introduce an auxiliary set of four Grassmann variables $\phi_\alpha$, and define an ``extended twistor" in ${\mathbb{CP}}^4$ by
\be
{\cal Z}^{{\cal I}}_i = \left( \begin{array}{c} Z^I_i \\ \overline{\phi \cdot \eta_i} \end{array} \right).
\ee
We also introduce the reference twistor
\be
{\cal Z}^{{\cal I}}_0 = \left( \begin{array}{c} 0 \\ 0 \\ 0 \\ 0 \\ 1 \end{array} \right)
\ee
which preserves the SL(4) symmetry acting on the bosonic $Z^I$.
It is natural to consider the (bosonic) volume of the 4-simplex whose faces are
${\cal Z}_a,{\cal Z}_b,{\cal Z}_c, {\cal Z}_d, {\cal Z}_e$:
\be
{\cal V} \left[{\cal Z}_a, \cdots, {\cal Z}_e\right] = \frac{1}{4!} \frac{\langle \langle{\cal Z}_a {\cal Z}_b {\cal Z}_c {\cal Z}_d {\cal Z}_e \rangle \rangle^4}{\langle \langle {\cal Z}_0 {\cal Z}_a {\cal Z}_b {\cal Z}_c {\cal Z}_d \rangle \rangle \langle \langle {\cal Z}_0 {\cal Z}_b {\cal Z}_c {\cal Z}_d {\cal Z}_e \rangle \rangle \cdots
\langle \langle {\cal Z}_0 {\cal Z}_e {\cal Z}_a {\cal Z}_b {\cal Z}_c \rangle \rangle}.
\ee
We use the notation $\langle \langle {\cal Z}_a {\cal Z}_b \cdots {\cal Z}_e \rangle \rangle$ to denote the contraction of the extended twistors with the 5-index $\epsilon_{{\cal I} {\cal J} {\cal K} {\cal L} {\cal M}}$ tensor, to distinguish it from the 4-brackets $\langle abcd \rangle$ used with the usual bosonic $Z_a^I$ twistors in ${\mathbb{CP}}^3$. We find trivially that
\be
{\cal V}\left[{\cal Z}_a, \cdots, {\cal Z}_e \right] = \frac{1}{4!} \frac{\left(\phi \cdot \eta_a \ab{bcde} + {\rm cyclic}\right)^4}{\ab{abcd} \ab{bcde} \cdots \ab{eabc}}.
\ee
This depends on $\phi$; to get a function of the $(Z_i, \eta_i)$ alone we simply integrate
over the $\phi_\alpha$. This directly yields the $R$-invariant
\be
\left[abcde\right] = \int d^4 \phi {\cal V}\left[{\cal Z}_a, \cdots, {\cal Z}_e\right].
\ee
Thus, in complete analogy with our warm-up example, we have associated the $[abcde]$ with (the $\phi$ integral of) the volume of a simplex in the ${\cal W}_{{\cal I}}$ space geometrically dual to ${\cal Z}^{{\cal I}}$ space, whose faces are ${\cal Z}_a,\cdots,{\cal Z}_e$. The algebraic properties of the $R$-invariants we have already discussed then precisely reflect the geometry of these simplices.

We can now give a nice definition for the NMHV amplitude polytope. Let us return to the BCFW/CSW expression
\be
M^{{\rm NMHV}}_n = \frac{1}{2}\sum_{i,j} \left[* \, i \, i+1 \, j \, j+1\right].
\ee
Mirroring our algebraic arguments from the introductory section, we can think of the $R$-invariants $\left[* \, i \, i+1 \, j \, j+1\right]$ as defining a 4-simplex in a (${\cal Z}$-space) ${\mathbb{CP}}^4$; the sum over all these tetrahedra defines a polytope $P_n$. $P_n$ is completely characterized by giving its boundary, which is
$\partial P_n = \sum_{i,j} \left[i \, i+1 \, j \, j+1 \right]$, showing that $P_n$ is actually independent of the point $*$.

We can think of $P_n$ as the ``square" of the Wilson-polygon $L_n$ in a natural way. Speaking slightly more generally, suppose we are given two ordered sets of points $X = (x_1, x_2, \cdots, x_n)$ and $Y = (y_1, y_2, \cdots, y_n)$, each of which defines an (in general non-planar) polygon loop in ${\mathbb{CP}}^4$.
We can form a 3-simplex $\left[x_i x_{i+1} y_j y_{j+1}\right]$ in ${\mathbb{CP}}^4$, by taking pairs of edges in $X$ and $Y$. Summing over all these 3-simplices defines a polyhedron $Q_n$, and it is easy to check that $\partial Q_n=0$ so $Q_n$ is a closed 3-volume. As such, we can we can write $Q_n = \partial (X \otimes Y)$, where $(X \otimes Y)$ is a 4-Polytope in ${\mathbb{CP}}^4$, one triangulation of which can be given as
$(X \otimes Y) = \sum_{i,j} \left[* \, i \, i+1 \, j \, j+1\right]$. Note  that we have used the ${\mathbb{CP}}^4$ structure in an essential way here,  in going from $Q_n$ being closed to being expressed as the boundary of a unique 4-dimensional object. Note also that the $\otimes$ operation behaves as an algebraic direct product in that it is linear in its two factors.
It is also interesting to note that, while the $X,Y$ polygons are in general non-planar, they nonetheless behave as plane polygons in this product, as reflected in the fact that $X \otimes Y$ satisfies 4-term tetrahedral identities separately in $X$ and $Y$.

With this definition, the polytope $P_n$ associated with the NMHV tree amplitude is related to the Wilson-Loop Polygon $L_n$ as
\be
P_n = \frac{1}{2} L_n \otimes L_n .
\ee
The NMHV amplitude is $\int d^4 \phi$ of the volume of the polytope $\widetilde{P}_n$ geometrically dual to $P_n$, which we can represent as a contour integral via
\be
M^{{\rm NMHV}}_n = 4!  \int d^4 \phi \int \limits_{\widetilde{P}_n} \frac{D^4 {\cal W}}{({\cal Z}_0 \cdot {\cal W})^5}.
\ee

In order to actually compute this volume, we need a triangulation of $\widetilde{P}_n$ in terms of elementary 4-simplices.
We may triangulate the polytope in any way we like. The BCFW representation of the amplitude is one particular choice, which yields the shortest expressions for the amplitude but has spurious poles.  The BCFW triangulation adds no new planes, but does add new vertices, and the spurious poles are associated with these ``spurious" vertices. The geometrically dual choice --- adding no new vertices but adding spurious planes --- will yield expressions for the amplitude that
allow us to expose manifest cyclicity and locality in a new way.

We do this by first triangulating each of the {\it faces} of $\widetilde{P}_n$. All the boundaries of $\widetilde{P}_n$ lie in the planes dual to the ${\cal Z}_j$; we denote the face contained in this plane by $F_{j,n}$. Conveniently, the faces are 3-polytopes, which will allow us to visualize them easily.
We can triangulate $F_{j,n} = \sum_\gamma T^\gamma_{j,n}$, where each of the $T_{j,n}^\gamma$ is a tetrahedron with 4 vertices. In order to triangulate $\widetilde{P}_n$, we introduce a reference ``suspension point" ${\cal W}_*$. With each tetrahedron $T^\gamma_{j,n}$, we associate a 4-simplex ${\cal T}^\gamma_{j,n}$ just by adding the point ${\cal W}_*$ to the 4 vertices of ${\cal T}^\gamma_{j,n}$. The sum over all these 4-simplices then gives a triangulation of $\widetilde{P}_n$ given by $\widetilde{P}_n = \sum_{j,\gamma} {\cal T}^\gamma_{j,n}$.

We have a natural choice for the ``suspension point" ${\cal W}_*$. Given that our choice of ${\cal Z}_0$ leaves the SL(4) acting on the usual bosonic momentum-twistors invariant, it is natural to choose ${\cal W}_*$ to also preserve this SL(4). Explicitly, we can choose ${\cal W}_* = (0,0,0,0,1)$. Finally, for a ``local" triangulation, we will choose to triangulate the faces only using the given ``physical" vertices $(i \, i\!+\!1 \, j \, j\!+\!1)$.

Following \cite{Hodges:2009hk}, let us get acquainted with the faces of $\tilde{P}_n$ by looking at $F_{2,n}$.
The vertices of $F_{2,n}$ are all the points of the form $(1 \, 2 \, k \, k\!+\!1)$ and $(2 \,3 \, l \, l\!+\!1)$. Two vertices $(2abc),(2xyz)$ are connected by an edge if the triples $(abc),(xyz)$ share two indices in common.
In the simplest case $n=5$, the face $F_{2,5}$ is just a tetrahedron with vertices $(1234),(1245),(2345),(2351)$. For 6 particles, $F_{2,6}$ has six vertices, and while e.g. $(2356)$ is connected by an edge to $(2345)$, there is no edge connecting $(2356)$ to $(1234)$.

It is very easy to recursively build $F_{2,n}$ systematically, starting from the tetrahedron for $F_{2,5}$.
While the vertices $(1234),(1245),(2345)$ occur in both $F_{2,5}$ and $F_{2,6}$, the vertex $(2351)$ occurring in $F_{2,5}$ is absent in $F_{2,6}$; conversely there are three new vertices $(2356),(1256)$ and $(2356)$ in $F_{2,6}$ not contained in $F_{2,5}$. Thus we can obtain $F_{2,6}$ by starting with $F_{2,5}$, ``chopping off" the vertex $(2351)$ and replacing it with the three new vertices, as shown below:
\be
\includegraphics[scale=.35]{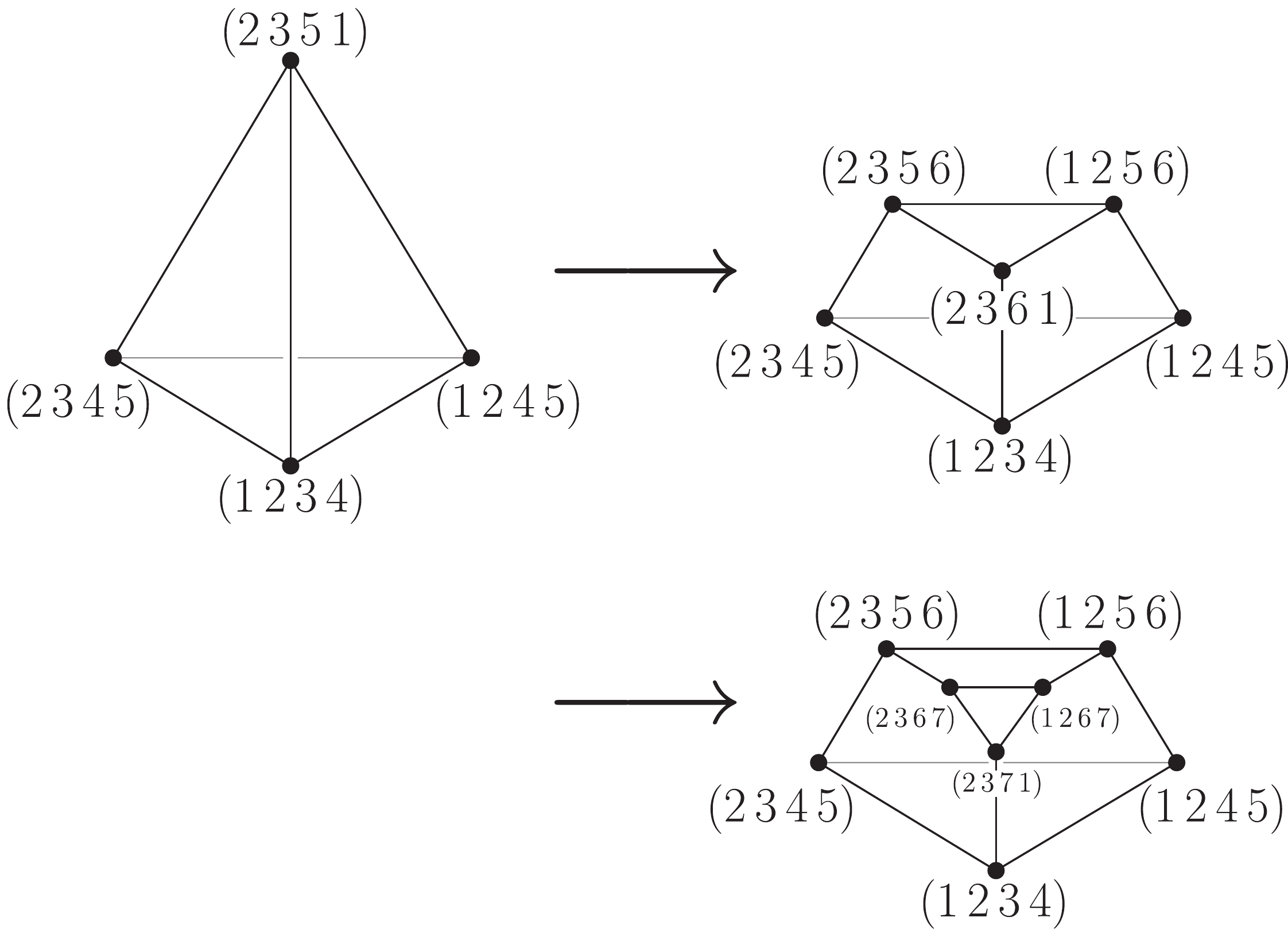} \nonumber
\ee
Similarly, we can go from the $F_{2,6}$ to $F_{2,7}$ by ``chopping off" the vertex $(2361)$, and continue in this way to define all the $F_{2,n}$.

We now wish to give a local triangulation of the $F_{2,n}$. Let's illustrate this in pictures with one local triangulation for the first non-trivial case of $F_{2,6}$:

\be
\includegraphics[scale=.35]{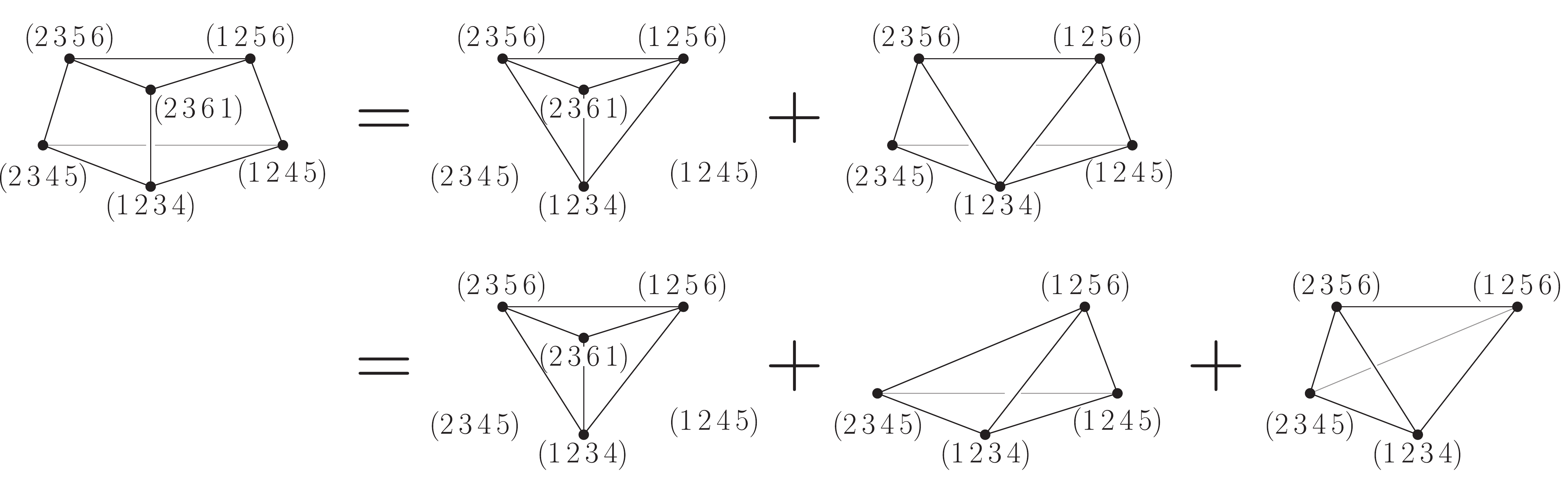} \nonumber
\ee
This ``local" triangulation is to be contrasted with a ``BCFW" triangulation, which would adds back the ``spurious" point $(2351)$, and represent the prism for $F_{3,6}$ as $F_{2,5}$ with the ``chopped off"
tetrahedron subtracted.

For general $n$, we can build $F_{2,n}$ from $F_{2,n-1}$ by ``chopping off" the vertex $(2\, 3\,  n\!-\!1 \,1)$ and adding the three vertices $(2\, 3 \, n \,1), (2 \, 3 \, n\!-\!1 \, n), (1 \, 2 \, n\!-\!1 \, n)$. This makes it natural to define
\be
F_{2,n} = G_n + T_n
\ee
where $T_n$ is the tetrahedron
\be
T_n = \raisebox{-1.5cm}{\includegraphics[scale=0.4]{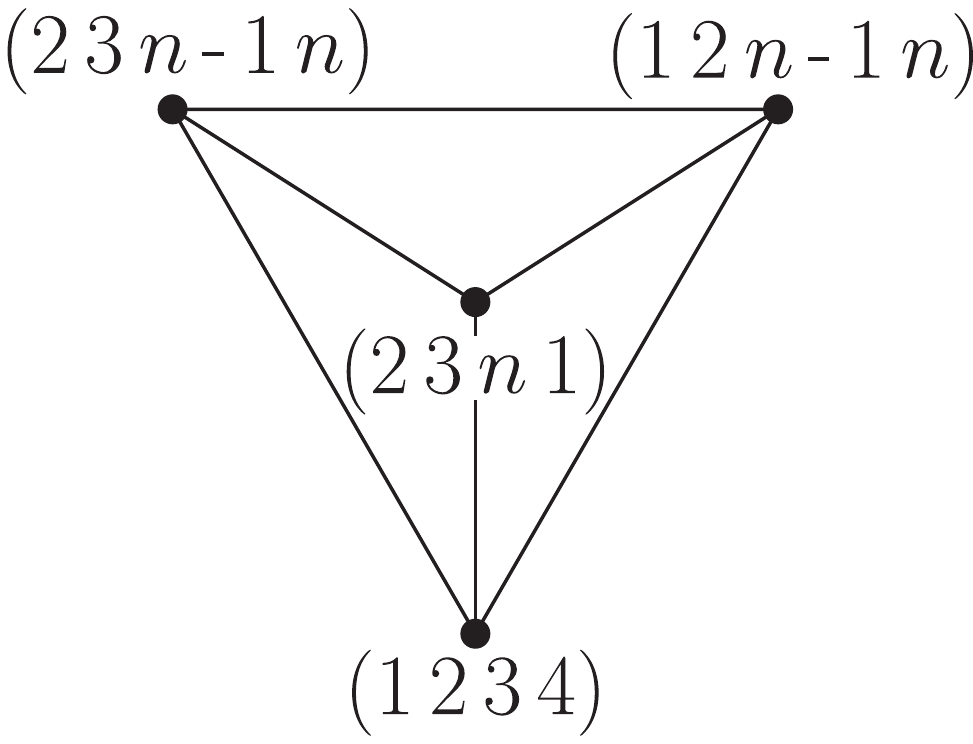}}
\ee
In going from $F_{2,n-1}$ to $F_{2,n}$, we just chop off the vertex $(2\, 3 \, n\!-\! 1\, 1)$ from $T_{n-1}$, thus we can write
\be
F_{2,n} = G_{n-1} + \raisebox{-1.5cm}{\includegraphics[scale=0.4]{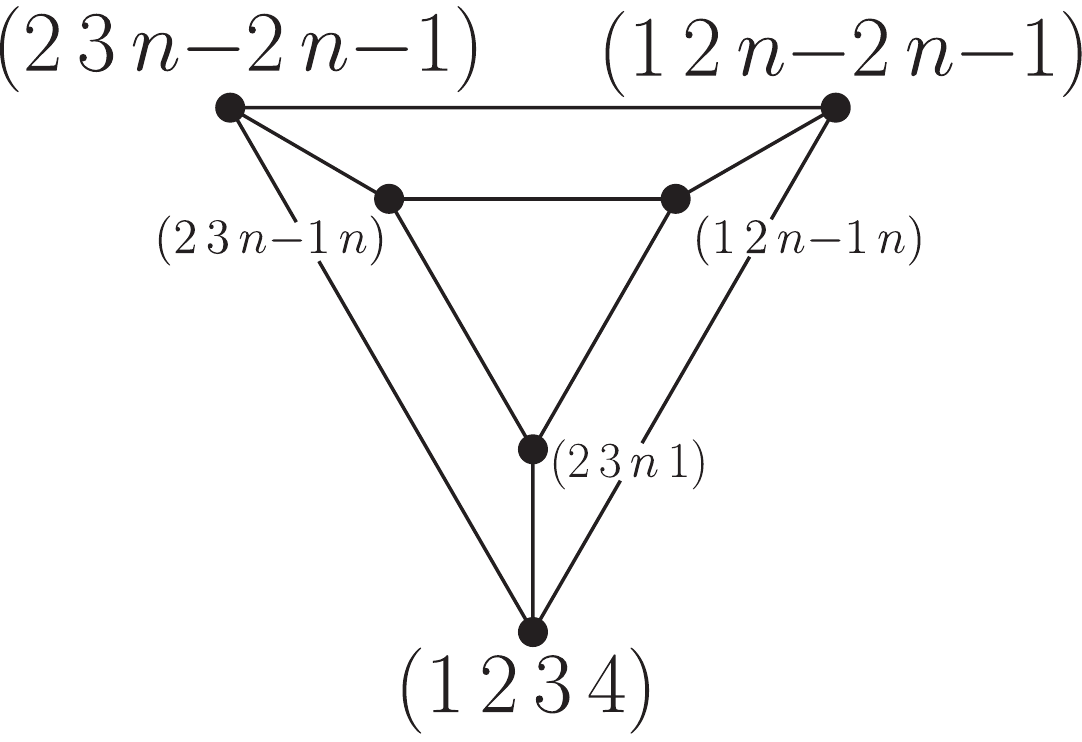}} = G_{n-1} + T_{n} +
\raisebox{-1.5cm}{\includegraphics[scale=0.4]{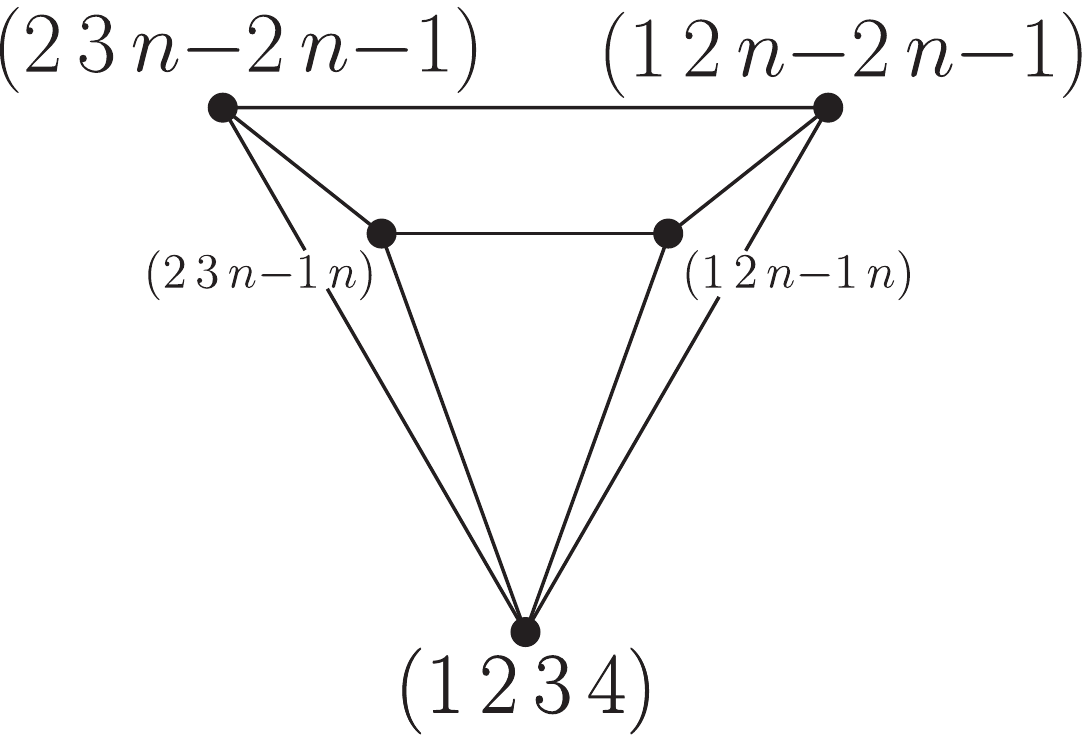}}
\ee
Since $F_{2,n} = G_{n} + T_{n}$, this gives us a recursive formula for the $G_n$:
\be
G_{n} = G_{n-1} + \!\!\!\!\!\! \raisebox{-1.5cm}{\includegraphics[scale=0.4]{fig9b.pdf}}\!\!\!\!\!\!\! = G_{n-1} + \!\!\!\!\!\! \raisebox{-1.5cm}{\includegraphics[scale=0.4]{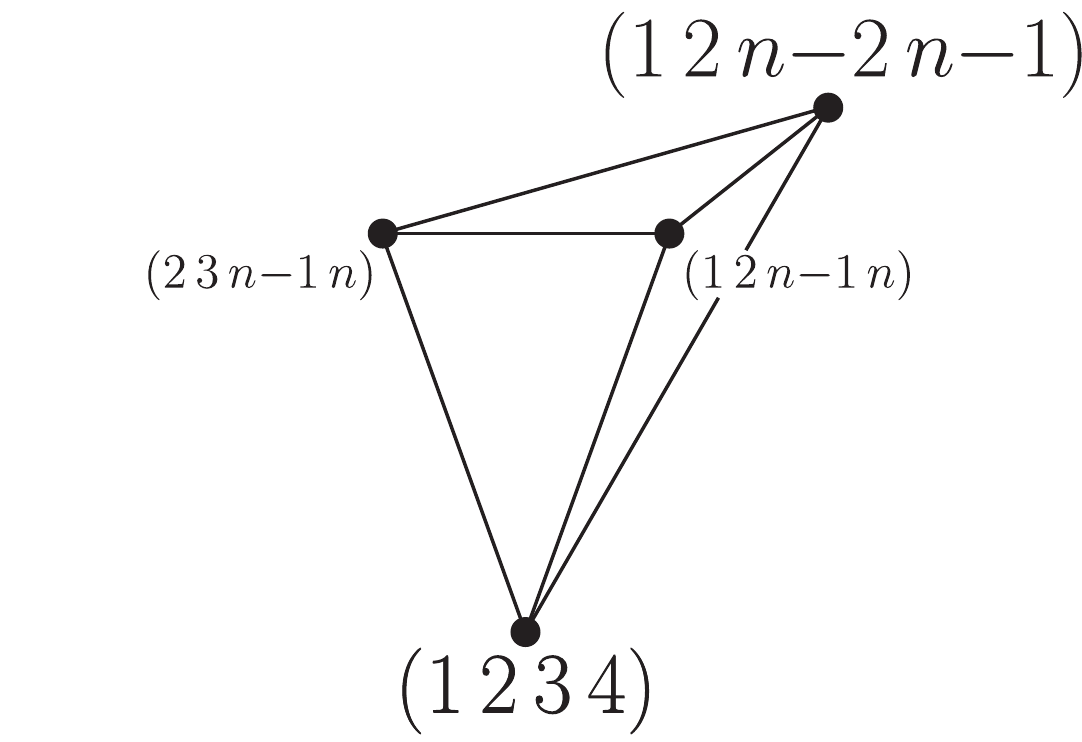}} + \raisebox{-1.5cm}{\includegraphics[scale=0.4]{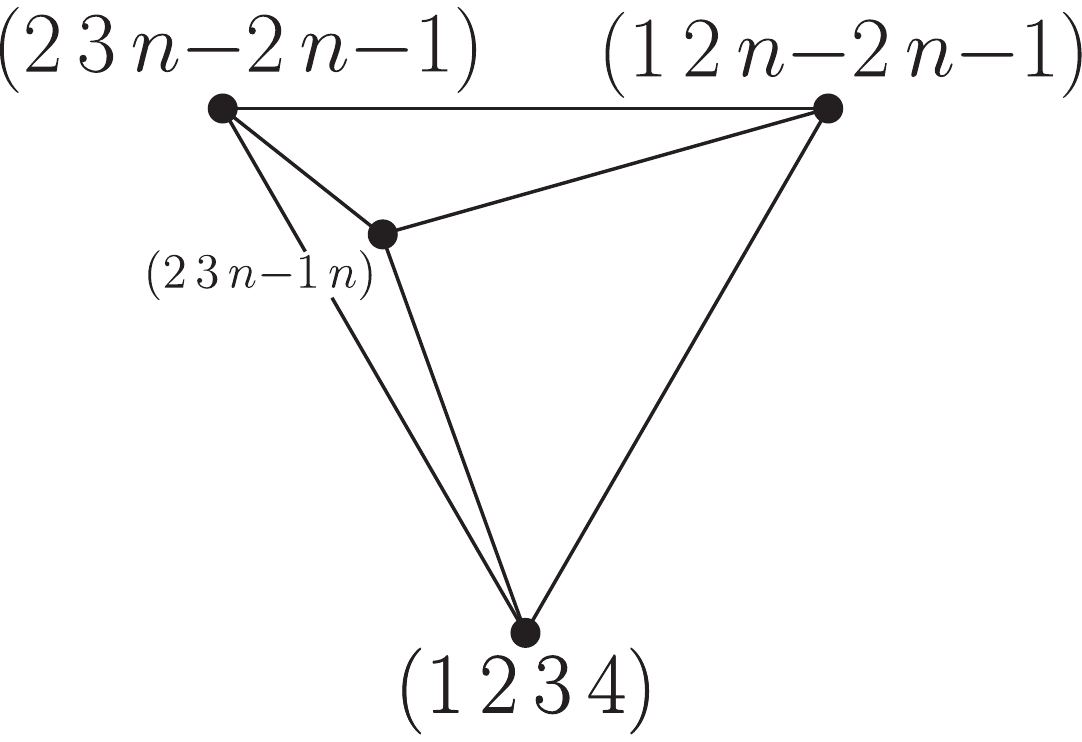}}\nonumber
\ee
Note that in the last step we chose one of two possible triangulations of the prism occurring in the first term. This is not fundamental, we have made this choice because it will lead to the a slightly simpler final form for the amplitude.

We can trivially solve for all the $G_n$. Doing this and adding $T_n$ we find for $F_{2,n}$
\be
F_{2,n} = \sum_i\left( \raisebox{-1.5cm}{\includegraphics[scale=0.4]{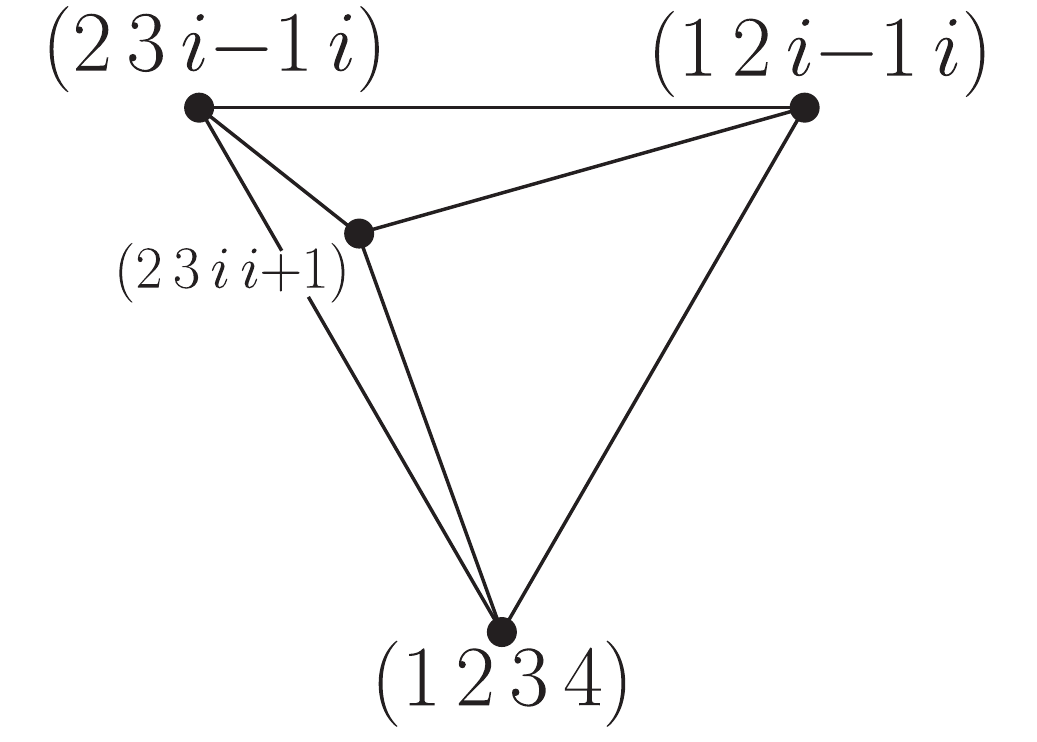}} + \raisebox{-1.5cm}{\includegraphics[scale=0.4]{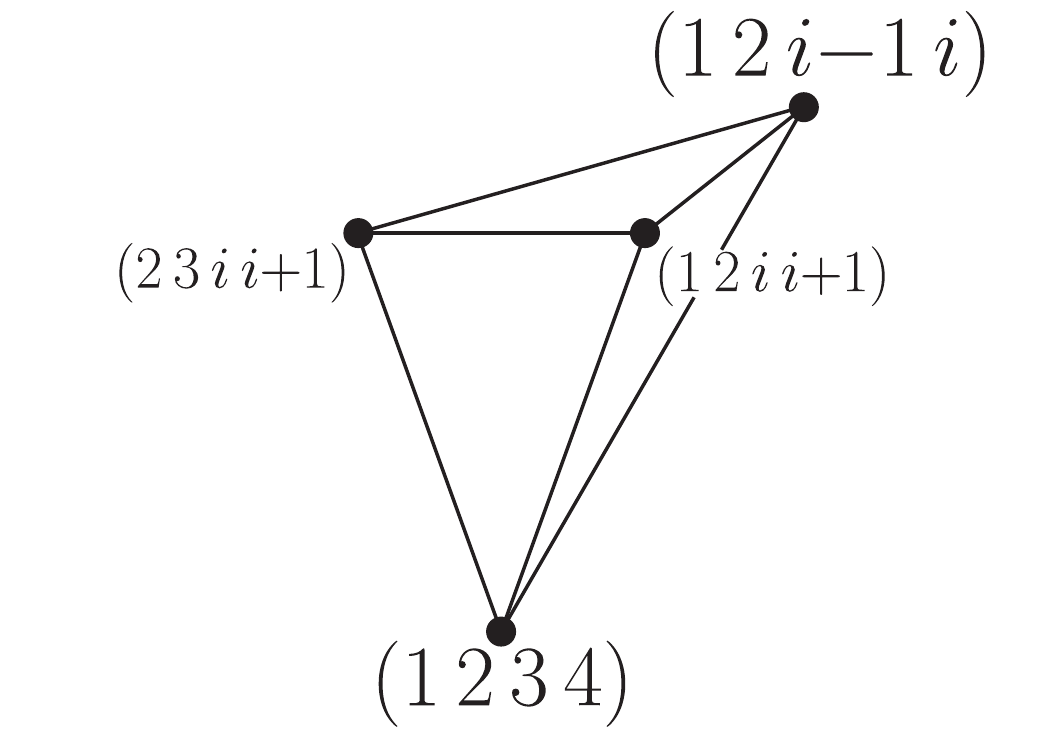}}\right)
\ee
note that the $T_n$ contribution is nicely represented by the term with $i = n$ in the first sum. Note also that we sum over all $i$ without worrying about any limits since any degenerate
configurations have vanishing volume. Actually it is easy to see that the number of non-degenerate terms is $2n-9$ since we start with one tetrahedron when $n=5$, and then each increase in $n$ needs one more chopping which generates two more terms.

Obviously this procedure works for any face $F_{j,n}$ just by cycling labels, and the final result for $F_{j,n}$ is
\be
\label{face}
F_{j,n} = \sum_i \left( \raisebox{-1.5cm}{\includegraphics[scale=0.4]{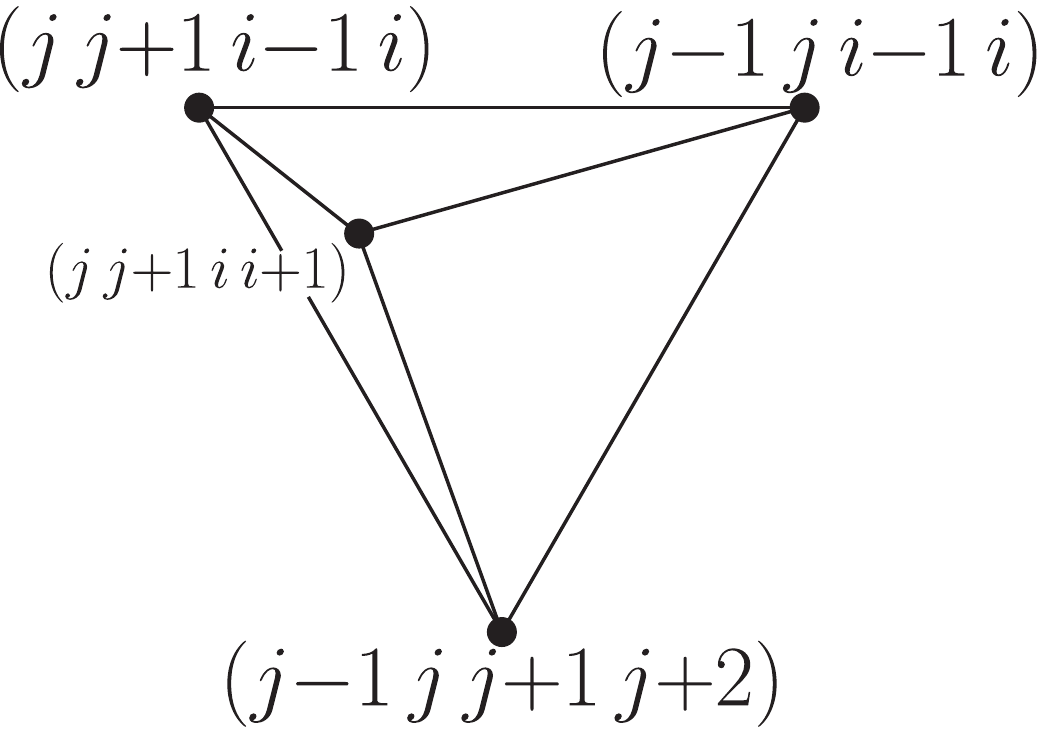}} \!\!\!\!\!\! + \raisebox{-1.5cm}{\includegraphics[scale=0.4]{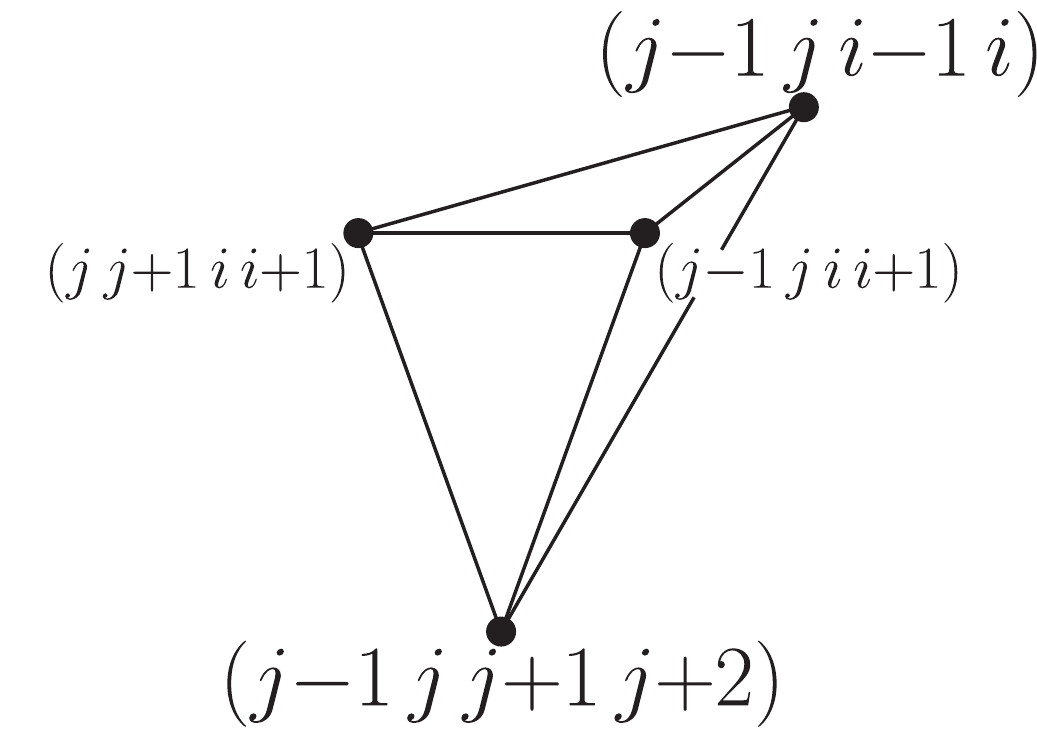}}\right) = \sum_{i;s = \pm 1}\!\!\!\!\!\!\! \raisebox{-1.5cm}{\includegraphics[scale=0.4]{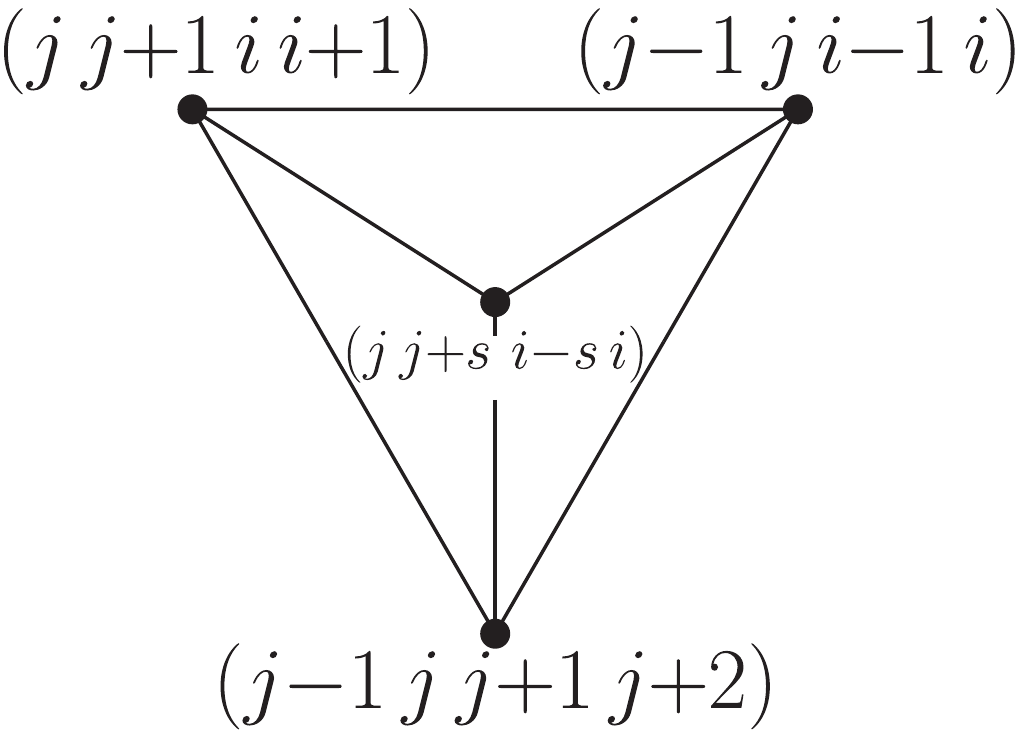}}
\ee

Now that we have the triangulation of the faces, we can easily compute the volume of the triangulations of the polytope itself, involving the addition of our suspension point ${\cal W}_* = (0,0,0,0,1)$. The volume associated with the tetrahedra appearing in \mbox{eqn. (\ref{face})} for the face $F_{j,n}$, is
\be
\frac{\langle \langle {\cal W}_* \, (j\!-\!1 \, j \, j\!+\!1 \, j\!+\!2) \, (j\!-\!1 \, j \, i\!-\!1\, i) \, (j \, j\!+\!1 \, i \, i\!+\!1)\, (j \, j\!+\!s \, i\!-\!s \, i) \rangle \rangle}{\ab{j\!-\!1 \, j \, j\!+\!1 \, j\!+\!2}
\ab{j\!-\!1 \, j \, i\!-\!1 \, i} \ab{j \, j\!+\!1 \, i \, i\!+\!1}
\ab{j \, j\!+\!s \, i\!-\!s \, i}}
\ee
which can easily be computed. There are 16  ${\cal Z}^{{\cal I}}$'s
in the numerator and a single dual twistor ${\cal W}_{* {\cal I}}$. Thus the 5-bracket expands to a sum of terms where one of the ${\cal Z}$'s is contracted with ${\cal W}_*$, and the remaining 15 are grouped into the product of three 5-brackets. Since ${\cal Z}_j$ occurs four times, for a non-zero result one of these ${\cal Z}_j$  must be contracted with ${\cal W}_*$. There is only one non-vanishing way of grouping the remaining ${\cal Z}$'s into 5-brackets, and we find this volume to be
\be
\frac{\phi \cdot \eta_j \langle \langle j\!-\!1 \, j \, j\!+\!1 \, j\!+\!2 \, i \rangle \rangle \langle \langle j\!-\!1 \, j \, j\!+\!1 \,
 i\!-\!s \, i\rangle \rangle \langle \langle j \, j\!+\!s \, i\!-\!1 \, i \, i\!+\!1 \rangle \rangle}{\ab{j\!-\!1 \, j \, j\!+\!1 \, j\!+\!2}
\ab{j\!-\!1 \, j \, i\!-\!1 \, i} \ab{j \, j\!+\!1 \, i \, i\!+\!1}
\ab{j \, j\!+\!s \, i\!-\!s \, i}}
\ee
the $\int d^4 \phi$ integration is trivially done to yield
\be
\frac{\langle \eta_j , \{j\!-\!1 \, j \, j\!+\!1 \, j+2 \, i\}, \{j\!-\!1 \, j \, j\!+\!1 \, i\!-\!s \, i\}, \{j \, j\!+\!s \,
i\!-\!1 \, i \, i\!+\!1\} \rangle}{\ab{j\!-\!1 \, j \, j\!+\!1 \, j\!+\!2}
\ab{j\!-\!1 \, j \, i\!-\!1 \, i} \ab{j \, j\!+\!1 \, i \, i\!+\!1}
\ab{j \, j\!+\!s \, i\!-\!s \, i}}.
\ee
Here, we have defined the Grassmann object
\be
\{abcde\} = \eta_a \ab{bcde} + \cdots + \eta_e \ab{abcd}
\ee
and the four-bracket in the numerator represents the contraction of SU(4)$_{\rm R}$ indices of the $\eta$'s.

We have thus found a manifestly cyclic and local formula for the NMHV tree amplitude
\be
\label{locnmhv}
M^{{\rm NMHV}}_n = \sum_{i,j;s=\pm 1}
\frac{\langle \eta_j , \{j\!-\!1 \, j \, j\!+\!1 \, j\!+\!2 \, i\}, \{j\!-\!1 \, j \, j\!+\!1 \, i\!-\!s \, i\}, \{j \, j\!+\!s \,
i\!-\!1 \, i \, i\!+\!1\} \rangle}{\ab{j\!-\!1 \, j \, j\!+\!1 \, j\!+\!2}
\ab{j\!-\!1 \, j \, i\!-\!1 \, i}, \ab{j \, j\!+\!1 \, i \, i\!+\!1}
\ab{j \, j\!+\!s \, i\!-\!s \, i}} .
\ee
This expression is amazingly simple, with $n(2n-9)$ non-vanishing terms. Here $(2n-9)$ is simply the number of tetrahedra in each face we already encountered in \mbox{eqn. (\ref{face})}. It also has another striking property: despite naturally being written as a function of the supersymmetric Grassmann $\eta$ variables, the individual terms in the sum are {\it not} invariant under supersymmetry transformation (to speak nothing of the Yangian symmetry). Indeed, the SUSY variation cancels only in a telescopic sum over all the terms.

There are other possible local triangulations of this polytope.  For instance, we can choose the ``suspension point" to be one of the vertices of the polytope $(k\,  k\!+\!1 \, j \, j\!+\!1)$. This is analogous to the choice ``$X = (n1)$''  for the reference bitwistor in the local form of the MHV 1-loop integrand, and gives a shorter formula with $(n-4)(2n - 9)$ terms, but at the  cost of losing manifest cyclic symmetry. These local forms can finally be compared with the BCFW expressions with $\frac{1}{2}(n-3)(n-4)$ terms, which contain $\sim \frac{1}{4}$ as many terms at large $n$, but don't make cyclicity or locality manifest.

It is also amusing to give a formula for standard helicity amplitudes; this only requires computing simple determinants as explained in e.g. \cite{ArkaniHamed:2009sx}. Consider in particular gluon amplitudes $A_n(1^- \, i^- \,  j^-)$ where particles $1,i,j$ have negative helicity and the rest have positive helicity. Up to the Parke-Taylor pre-factor, the result is precisely given by the above formula for $M^{{\rm NMHV}}_n$, with the $\eta_k$ replaced by a particular function of the spinor-helicity variables. Explicitly, we find that
\be
\label{comp}
A_n(1^- \, i^- \, j^-) = \frac{1}{\ab{12} \ab{23} \cdots \ab{n1}} M^{{\rm NMHV}}_n \left(\eta_k \to \begin{array}{ccc} \ab{ij} \ab{k1} & {\rm for} & 1< k \leq i \\ \ab{kj} \ab{i1} & {\rm for} & i < k \leq j \\ 0 & {\rm otherwise} & \end{array} \right).
\ee

The split helicity amplitudes are particularly easy to extract for all $n$:
\be
A_n(1^- \, 2^- \, 3^-) = \frac{\ab{12}^4 \ab{23}^4}{\ab{12} \ab{23} \cdots \ab{n1}}
 \sum_{i;s=\pm 1}
\frac{\langle 134\, i \rangle \langle 13 \, i\!-\!s \, i \rangle \langle 2\!+\!s \,
i\!-\!1 \, i \, i\!+\!1 \rangle}{\ab{1234}
\ab{12 \, i\!-\!1 \, i} \ab{23 \, i \, i\!+\!1}
\ab{2 \, 2\!+\!s \, i\!-\!s \, i}}.
\ee
Using  $\ab{i\!-\!1 \, i \, j \, j\!+\!1} = \ab{i-1 \, i} \ab{j \, j+1} (p_i + \cdots + p_j)^2$, the poles are directly functions of spinor-helicity variables and take the usual form of Feynman propagators.  For $n=6$, this expression is equivalent to a form derived long ago using the Berends-Giele recursion relations
\cite{Berends:1987me}; we now see that this formula and all its variant forms flow from the single formula, \mbox{eqn. (\ref{comp})}, which also generalizes to all helicity configurations and all $n$.

We conclude our discussion of NMHV amplitudes by remarking that the use of a  {\it bosonic} ${\mathbb{CP}}^4$ space to describe supersymmetric amplitudes is quite striking. One might have expected supersymmetric amplitudes to be expressed as an integral over ${\mathbb{CP}}^{3|4}$, and indeed the $R$-invariants have a beautiful interpretation  as the super-volume of a super-polytope \cite{Hodges:2009hk} in ${\mathbb{CP}}^{3|4}$. This form is also very closely related to the momentum-twistor Grassmannian formula \cite{Mason:2009qx}. The non-linear way in which ${\cal Z}_i,{\cal Z}_0$ package the supersymmetric information of the theory into only a single extra dimension is more novel and  interesting, and made the local triangulation leading to \mbox{eqn. (\ref{locnmhv})} possible. We expect that further generalizations of this idea are needed for higher N$^k$MHV amplitudes.

\section{Discussion}

Many of the advances in our understanding of perturbative scattering amplitudes in the last five years were driven by the discovery of the CSW and BCFW recursion relations for tree amplitudes. The ability to analytically compute all tree amplitudes enabled the generation of a huge amount of ``data" about the theory, which exposed a number of new, remarkable and deeply interwoven mathematical structures underlying the physics. Amongst other things, these insights stimulated the generalization of the early methods to all loop orders, making a
more incisive exploration of the structure of the theory possible. In this note we have continued the exploration of one of the beautiful structures uncovered in this period.

The polytope picture is clearly intimately related to the Grassmannian formula in momentum twistor space, giving  a lovely geometric understanding of the  additive structures appearing in the amplitudes, which are understood more algebraically as a consequence of residue theorems in the Grassmannian formalism. 

The Grassmannian picture extends to all amplitudes and loop orders, giving expressions that are term-by-term manifestly Yangian-invariant. There is clearly a beautiful algebraic structure at work in governing the properties of  Grassmannian residues and residue theorems, guaranteeing the emergence of physical properties such as cyclic invariance, locality and unitarity. While we have not yet extended the polytope picture to these more general amplitudes, there are strong reasons to suspect this must be possible, and we expect that such an extension would give a more geometric understanding of these algebraic structures.

However, even in the baby examples we have studied in this note, it is clear that  the polytope picture does much more than simply geometrize the understanding of relations between Yangian invariants! While one simple class of polytope triangulations do indeed provide such an understanding, the even more natural class of triangulations we examined here have opened the door to a completely new set of objects and ideas, far removed from their BCFW/CSW origins.  The existence of such strikingly simple and manifestly local forms for the scattering amplitudes is a real surprise. Indeed the tremendous complexity of standard Feynman diagram calculations is directly related to making locality manifest, while the tremendous advantages of BCFW  seemed inexorably tied to the appearance of spurious poles!

This  strongly suggests a new set of principles at play. It is tempting to speculate that these principles will be closely connected to a more physical ``spin-chain" picture for scattering amplitudes. Superficially, the new expressions for the amplitudes we have found certainly {\it look} more closely related to an underlying spin-chain, not only on account of their manifest cyclicity, but also because of the suggestive way some of the symmetries are realized. It is also refreshing to move somewhat away from dealing with objects that are manifestly supersymmetric/Yangian invariant, particularly keeping in mind the eventual goal of understanding non-supersymmetric theories!

There is clearly some remarkable geometry behind these polytope formulas. It is particularly striking that in both of the examples we studied, the Wilson-Loop behaves as if it were a plane polygon, with additive identities like those of the triangles in ${\mathbb{CP}}^2$ explained in our warm-up example.

Finally, the polytope picture also strongly inspired the search for and discovery of the amazingly simple local expressions for multi-loop integrands reported in \cite{local}. These expressions are far simpler than their BCFW counterparts, and clearly beg for a much deeper understanding. We hope to see significant progress on these questions in 2011.

\section*{Acknowledgements}
We thank Lionel Mason, David Skinner, and Simon Caron-Huot for many inspiring discussions. A.H. would like to thank the Institute for Advanced Study and Perimeter Institute for their generous support in 2010. N.A.-H. is supported by the DOE under grant DE-FG02-
91ER40654, F.C. was supported in part by the NSERC of Canada, MEDT of Ontario and by The
Ambrose Monell Foundation. J.T. is supported by the U.S. Department of State through a Fulbright Science and Technology Award.

\end{document}